\documentclass[3p,times,procedia]{elsarticle}
\usepackage{nupha_ecrc}

\volume{00}

\firstpage{1}

\journalname{Nuclear Physics A}

\runauth{}

\jid{nupha}

\jnltitlelogo{Nuclear Physics A}

\usepackage{amssymb}

\usepackage{lineno}

\usepackage[figuresright]{rotating}

\begin{document}

\begin{frontmatter}



\dochead{XXVIIth International Conference on Ultrarelativistic Nucleus-Nucleus Collisions\\ (Quark Matter 2018)}

\title{Highlights from the STAR experiment}


\author{Zhenyu Ye (for the STAR collaboration)\footnote{A list of members of the STAR Collaboration and acknowledgements can be found at the end of this issue.}}

\address{University of Illinois at Chicago, 845 W. Taylor St., Chicago, IL 60607}

\begin{abstract}
The STAR experiment at the Relativistic Heavy Ion Collider (RHIC) studies properties and phase transitions of nuclear matter in various nucleus-nucleus collisions at center-of-mass energies per nucleon collision $\sqrt{s_{NN}}=7.7$-200 GeV. With a fixed target made of gold foils installed inside the beam pipe, STAR also starts to explore high baryon density regime ($\mu_B\approx420$-720 MeV) at $\sqrt{s_{NN}}=3.0$-7.7 GeV. A few selected results reported by the STAR collaboration at the Quark Matter 2018 (QM2018) conference are described in these proceedings. 
\end{abstract}


\end{frontmatter}


\section{Introduction}
The STAR experiment at RHIC is equipped with a Time Projection Chamber (TPC), a Time of Flight (TOF) detector, and a Barrel ElectroMagentic Calorimeter (BEMC) for particle detection and identification with full azimuthal coverage at mid-rapidity ($|\eta|<1$). At forward and backward rapidities, Beam-Beam Counters (BBC), Vertex Position Detectors (VPD) and Zero-Degree Calorimeters (ZDC) are used to trigger on minimum-bias collisions and for event plane reconstruction. A silicon vertex detector, the Heavy Flavor Tracker (HFT), was installed in STAR and took data in 2014-2016. It provided excellent vertex position and track pointing resolutions, allowing a detailed study of open heavy flavor hadron production at RHIC. Another detector made of Multi-gap Resistive Plate Chambers, the Muon Telescope Detector (MTD), has been fully installed into STAR since 2014, covering about 45\% of azimuth in the pseudo-rapidity range of $|\eta|<0.6$. It enables triggering and identification for muons with $p_T>1$ GeV/$c$, and thus opens the door to quarkonium measurements in the di-muon channel. In 2016, an electromagnetic calorimeter placed at $2.5<\eta<4$, the Forward Meson Spectrometer (FMS), participated in data taking in Au+Au collisions, allowing longitudinal flow decorrelation measurements at RHIC. These detectors with large acceptance and excellent performance, combined with the versatility of RHIC to collide different nuclei at a wide range of energies, put STAR in a unique position to study the properties of the Quark-Gluon Plasma (QGP) and phase transitions of nuclear matter. STAR contributed 18 oral presentations in parallel sessions and about 30 posters at the QM2018 conference. Some of these results are described below. Results not discussed here due to space limitation include jet-medium interaction \cite{jet}, event-plane dependence of di-hadron correlations with event-shape engineering \cite{dihadron}, $\phi$ and $K^*$ spin alignment \cite{spinalignment}, cumulants of net-charge and net-particle distributions \cite{cumulant}, flow and flow fluctuations in different collisions systems \cite{flow}, collectivity in small systems  \cite{small}, femtoscopy \cite{femtoscopy},  and fixed-target program \cite{FXT}. 

\section{Heavy Flavor Measurements}
\begin{figure}[b!]
\vspace{-0.2cm}
\center
\includegraphics[width=0.5\textwidth]{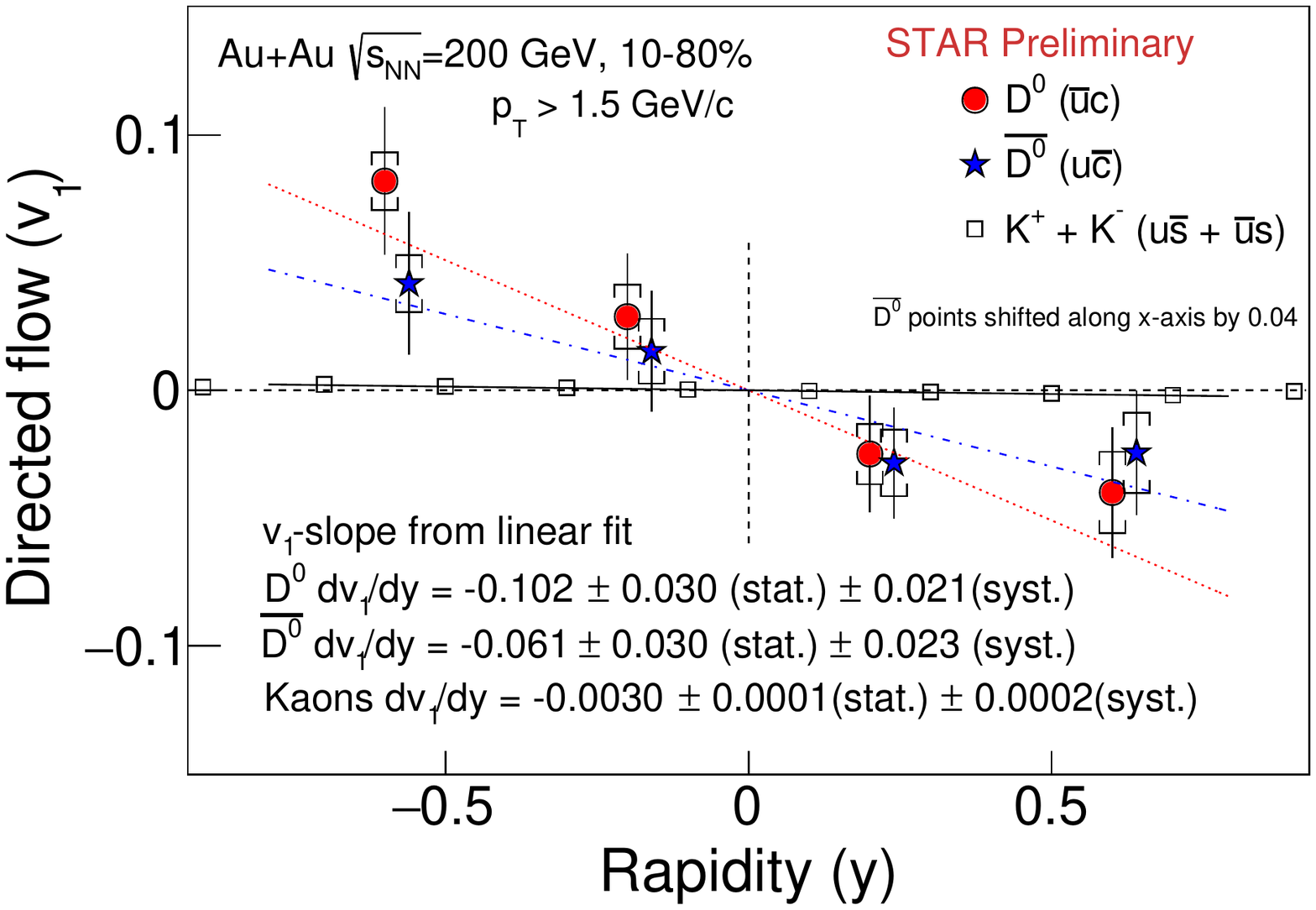}
\vspace{-0.5cm}
\caption{Directed flow $v_1$ for $D^0$ (circles), $\bar{D}^0$ (stars) and $K^\pm$ (boxes) as a function of rapidity $y$ in 10-80\% Au+Au collisions at $\sqrt{s_{NN}}=200$ GeV. Vertical bars (brackets) represent statistical (systematic) uncertainties. Also shown are linear fit results to these data.}
\label{fig:D0v1}

\includegraphics[width=0.8\textwidth]{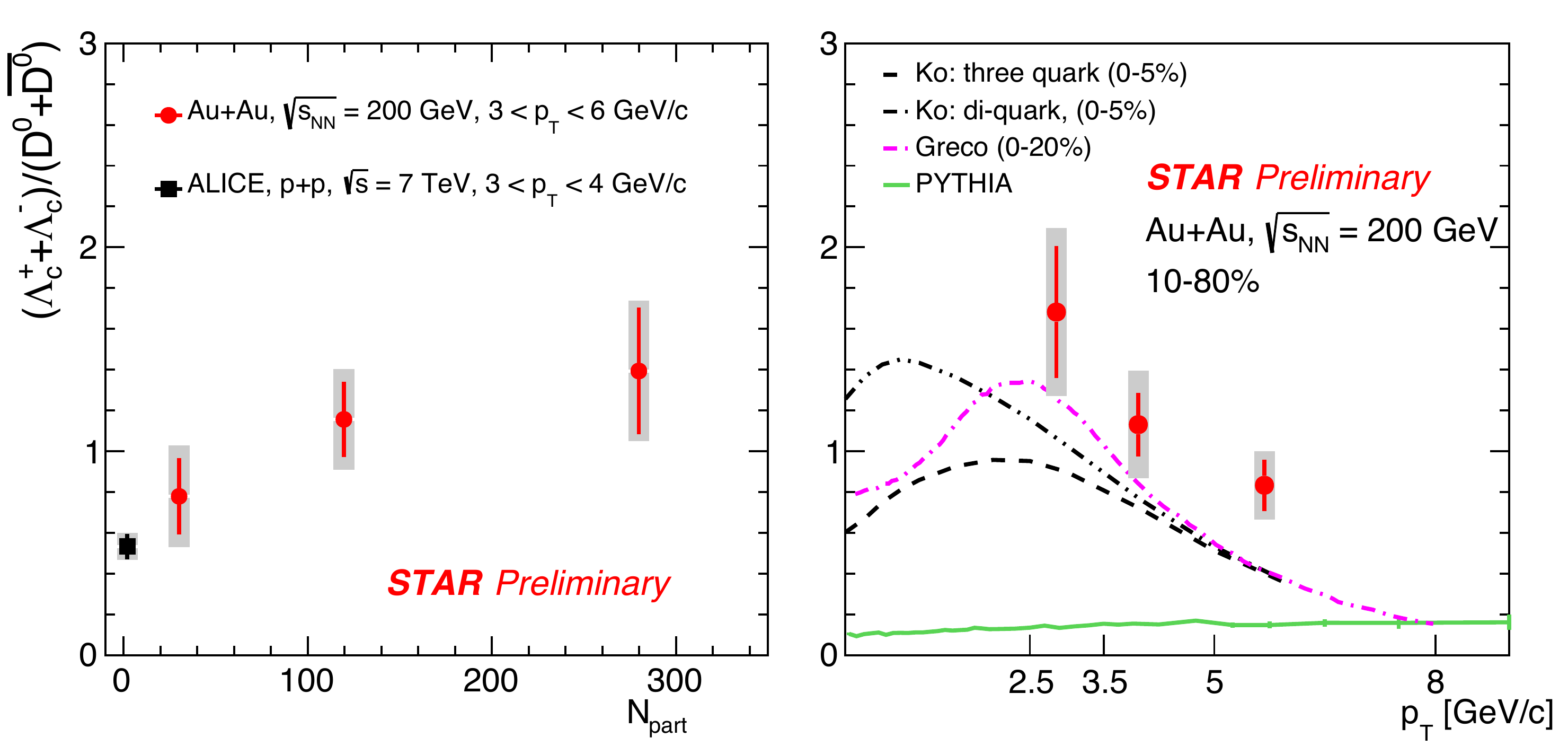}
\vspace{-0.3cm}
\caption{The $\Lambda_c/D^0$ ratio in Au+Au collisions at $\sqrt{s_{NN}}=200$ GeV as a function of centrality (left) and $p_T$ (right). Vertical bars (shaded areas) represent statistical (systematic) uncertainties. Also shown are the ratio in p+p collisions at $\sqrt{s}=7$ TeV measured by ALICE (black box), PYTHIA prediction (solid curve), and theoretical model calculations including charm quark coalescence (other curves).}
\label{fig:Lc}
\end{figure}

It has been suggested that the magnitude of the rapidity-odd directed flow $v_1$ for open charm hadrons is sensitive to the tilted shape of the initial source and to the charm quark drag coefficient in the QGP, while the difference in $v_1$ between charm and anti-charm hadrons is sensitive to the electromagnetic field in heavy-ion collisions \cite{D0v1Theory}. Thanks to the HFT, STAR has found the first evidence of a non-zero $v_1$ for open charm hadrons in heavy-ion collisions. The results are for Au+Au collisions at $\sqrt{s_{NN}}=200$ GeV, shown in Fig.~\ref{fig:D0v1} \cite{D0v1STAR}. The slope of $v_1$ with respect to $y$, averaged for the $D^0$ and $\bar{D}^0$ mesons, $dv_1/dy=-0.081\pm0.021(\mathrm{stat})\pm0.017(\mathrm{syst})$, is significantly larger than that of charged kaons. The $dv_1/dy$ values for the $D^0$ and $\bar{D}^0$ mesons are consistent with each other within uncertainties. 

The $\Lambda_c$ is the lightest baryon containing a charm quark. The $\Lambda_c$ yield at intermediate transverse momentum ($p_T$) is expected to be enhanced in heavy-ion collisions with respect to p+p collisions if charm quarks also hadronize via coalescence similarly to light flavor quarks. STAR presented the first result on the $\Lambda_c$ production in heavy-ion collisions at the QM2017 conference, obtained from the 2014 data with the HFT in Au+Au collisions at $\sqrt{s_{NN}}=200$ GeV. By adding the 2016 Au+Au data, and using a supervised machine-learning method based on boosted decision trees, the $\Lambda_c$ signal significance is increased by a factor of two as compared to the previous study. The $\Lambda_c/D^0$ yield ratio is determined as a function of centrality and $p_T$ and shown in Fig.~\ref{fig:Lc} \cite{STARLc}. It is found that the $\Lambda_c/D^0$ ratio increases towards more central Au+Au collisions, and the ratio in peripheral Au+Au collisions is compatible with that in p+p collisions at $\sqrt{s}=7$ TeV measured by ALICE \cite{ALICELc}. The ratio, when plotted as a function of $p_T$ and integrated over the 10-80\% centrality interval, is much closer to theoretical calculations considering charm quark coalescence \cite{TheoryLc} than the PYTHIA prediction in p+p collisions \cite{PYTHIA}. 

\begin{figure}[b!]
\includegraphics[width=0.5\textwidth]{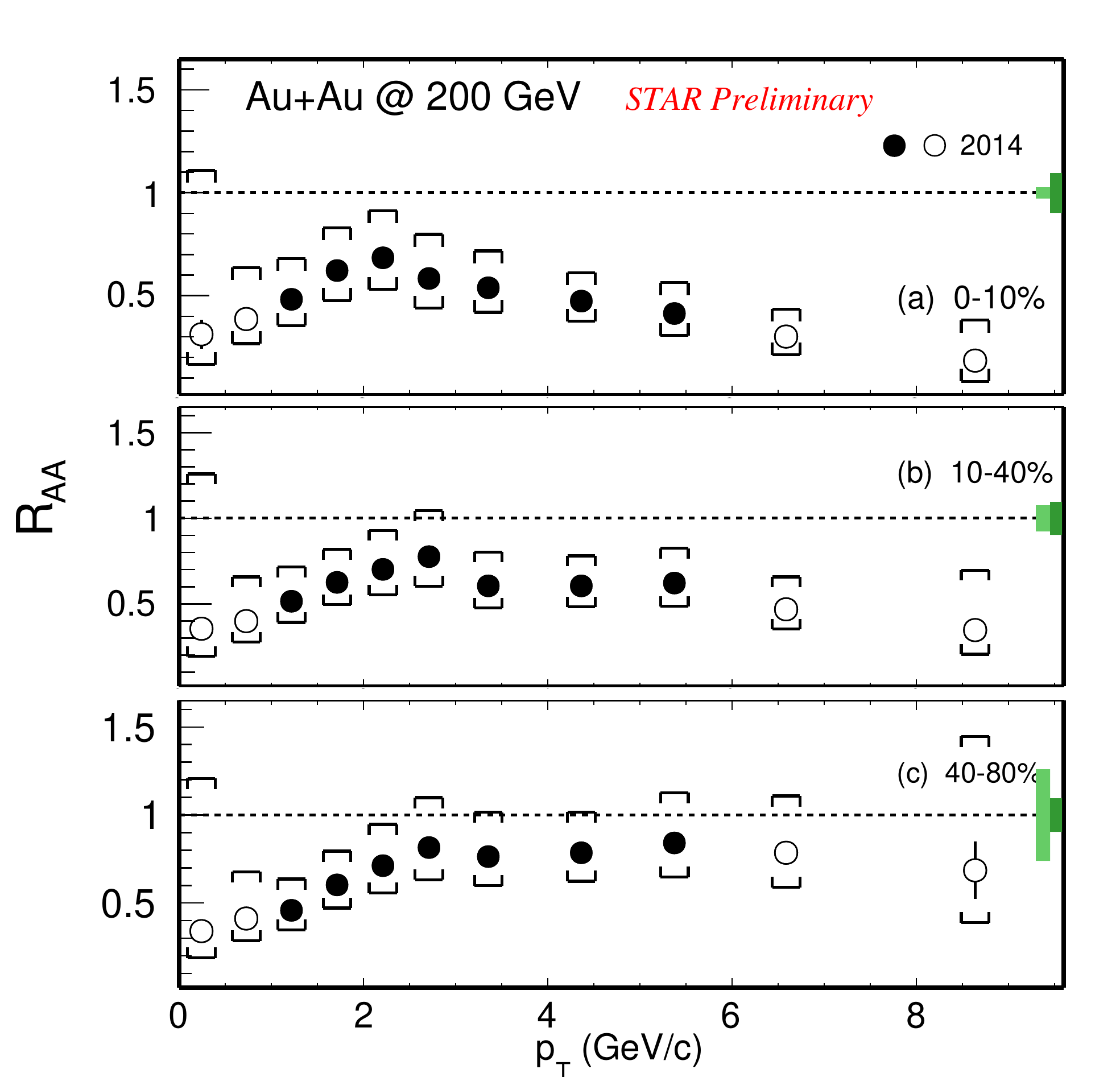}
\includegraphics[width=0.5\textwidth]{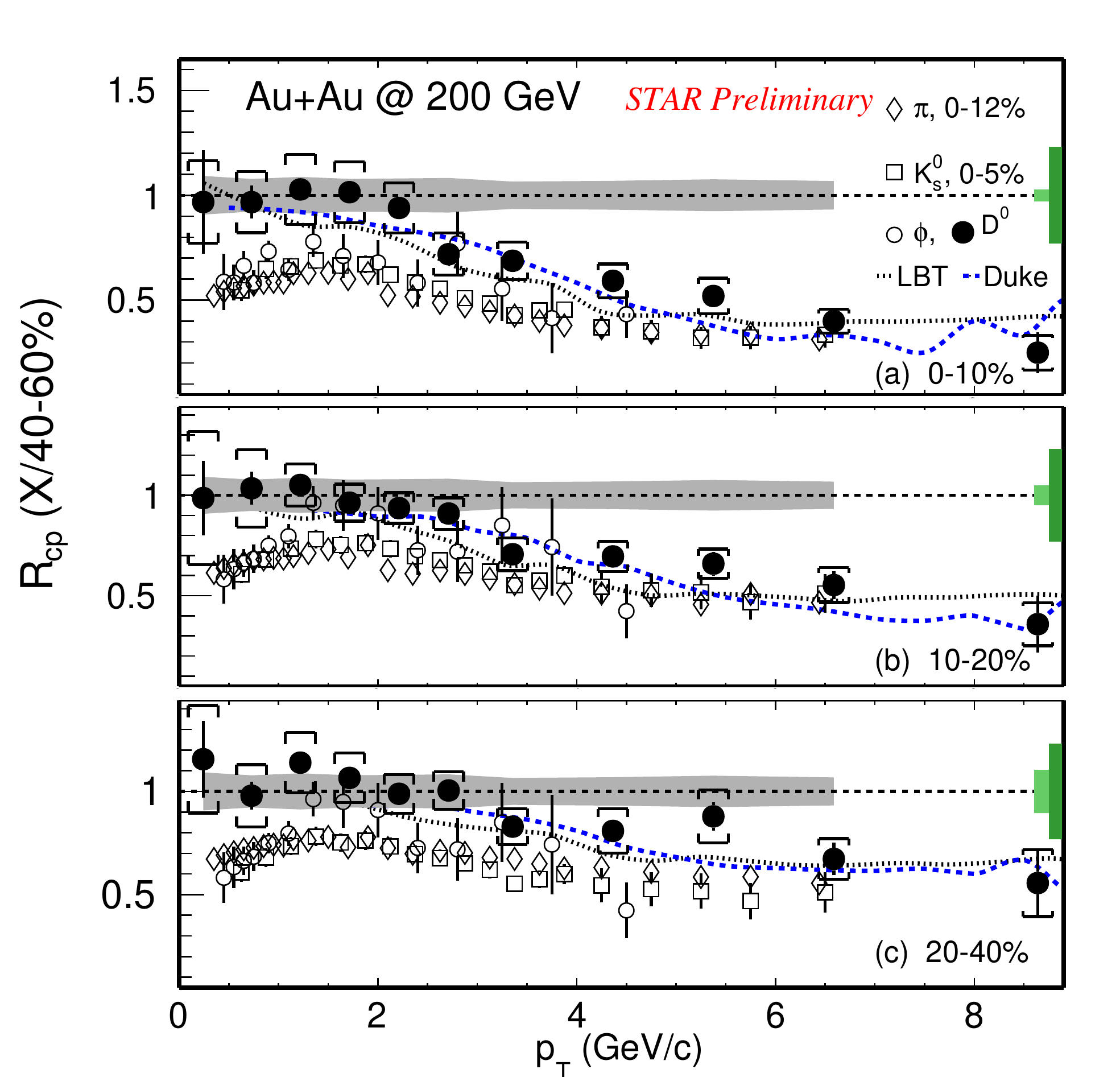}
\vspace{-0.8cm}
\caption{The $D^0$ nuclear modification factors $R_{AA}$ (left) and $R_{CP}$ (right) as a function of $p_T$ in different centrality intervals in Au+Au collisions at $\sqrt{s_{NN}}=200$ GeV. Vertical bars (brackets) represent statistical (systematic) uncertainties. The green boxes around unity at the right hand side of the panels are the Glauber model calculation uncertainties for $\left<N_{coll}\right>$, while the grey shaded areas around unity in the right figure are the uncertainties associated with vertex resolution correction. Also shown in the right panel are $R_{CP}$ for light flavor hadrons (open points) and theoretical model calculations for $D^0$ (dotted curves).}
\label{fig:D0RAA}

\vspace{0.3cm}
\includegraphics[width=0.5\textwidth]{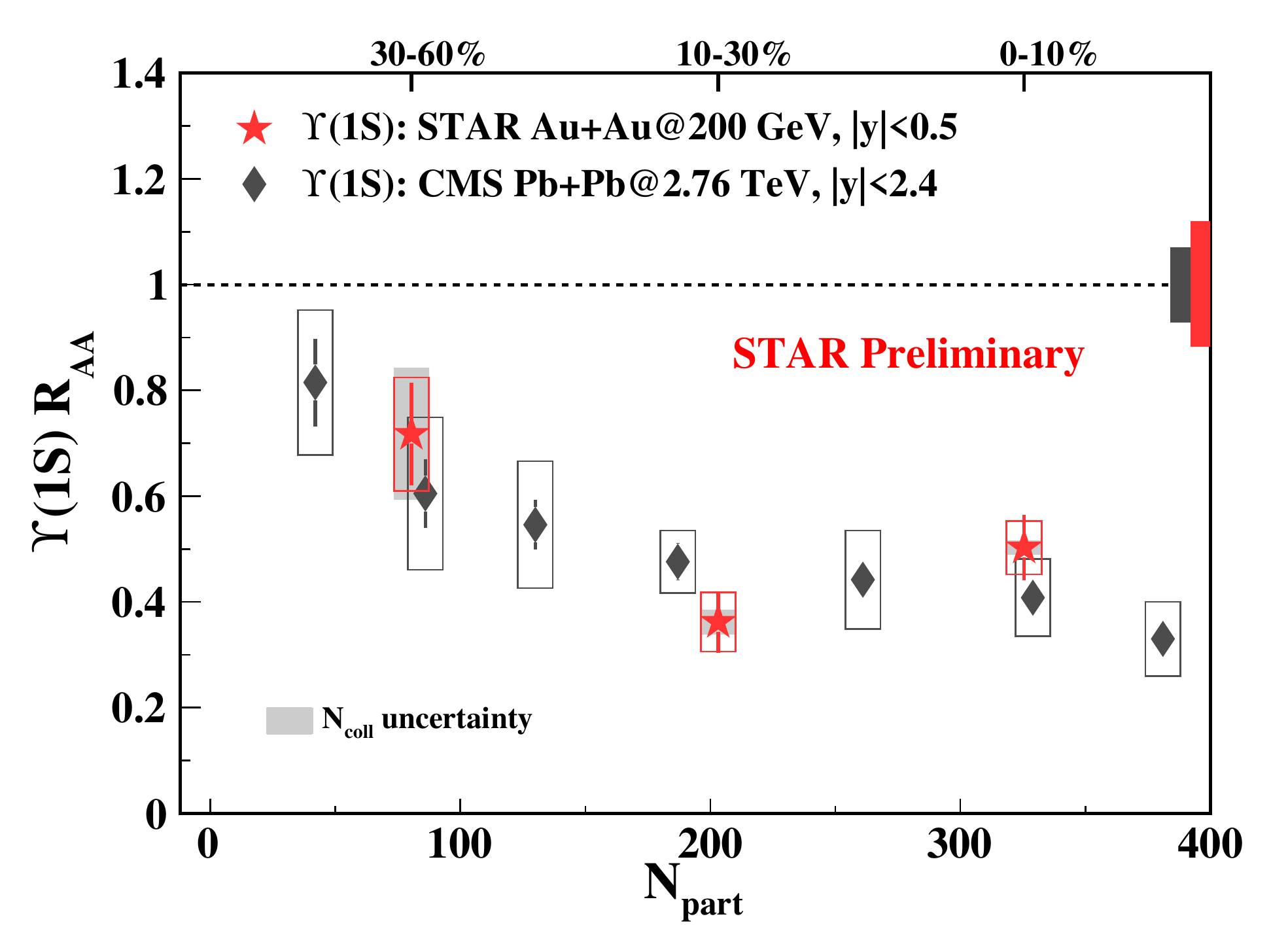}
\includegraphics[width=0.5\textwidth]{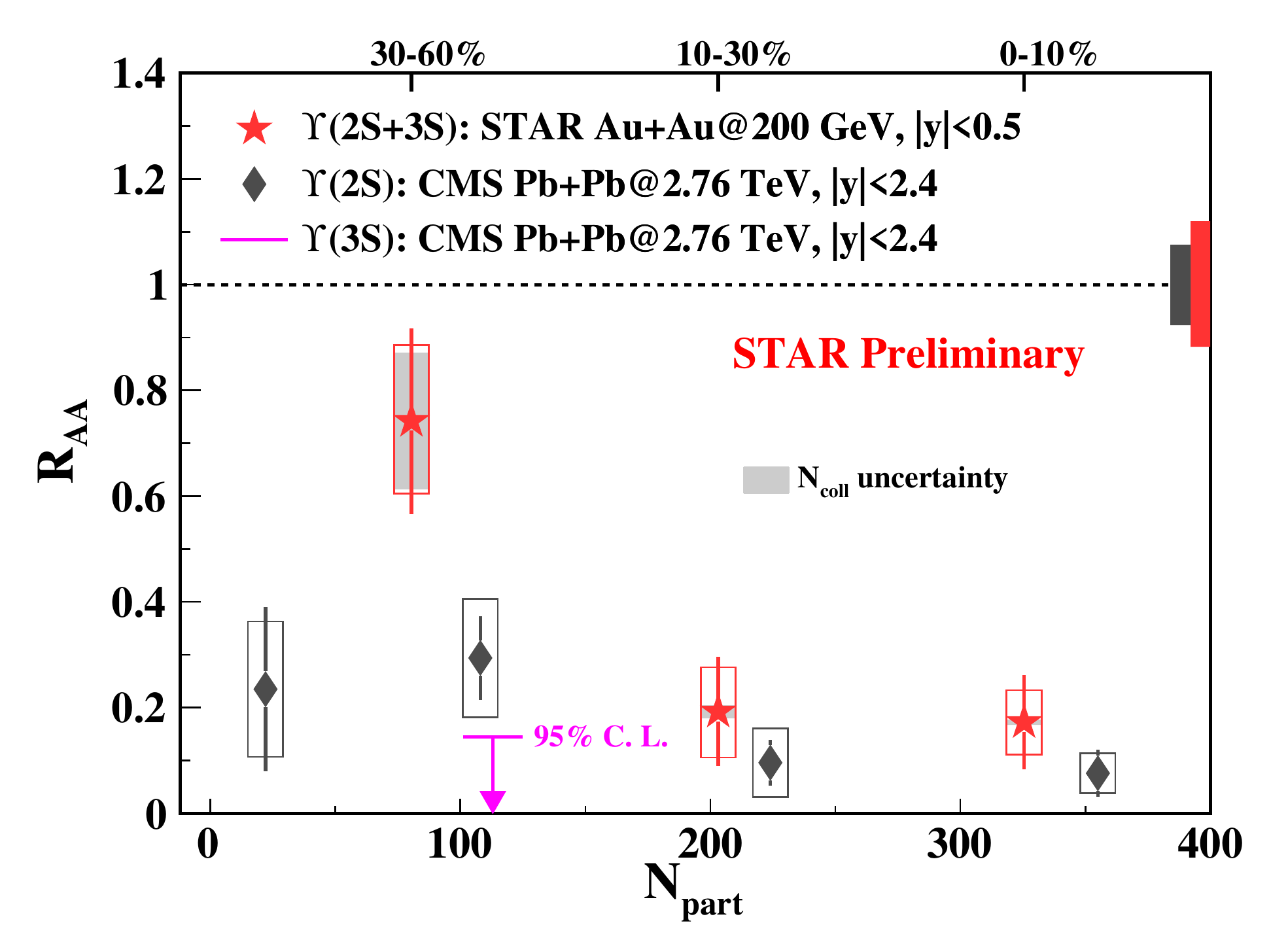}
\vspace{-0.8cm}
\caption{The $\Upsilon$ nuclear modification factor $R_{AA}$ for the ground (left) and excited (right) states integrated over $p_T$ as a function of the number of participating nucleons $N_{part}$. Vertical bars (open boxes) represent statistical (systematic) uncertainties, while the shaded gray area the Glauber model calculation uncertainty for the average number of binary nucleon-nucleon collisions $\left<N_{coll}\right>$. Filled boxes around unity are the global uncertainties dominated by that from the $p+p$ reference. Also shown are the results in Pb+Pb collisions at $\sqrt{s_{NN}}=2.76$ TeV from the CMS experiment \cite{UpsilonRAACMS}.}
\label{fig:Upsilon}
\end{figure}

Charm quark interactions with the QGP can also be studied through nuclear modification factors. The $R_{AA}$ and $R_{CP}$ for the $D^0$ meson are determined from the 2014 HFT data as a function of $p_T$ in different centrality intervals, as shown in Fig.~\ref{fig:D0RAA} \cite{STARLc}. A significant suppression of the $D^0$ yield at low $p_T$, which might come from cold nuclear matter effects and radial flow, is observed with little centrality dependence. A suppression is also observed at high $p_T$ in the most central collisions, which can be explained by energy loss of charm quarks when traversing the QGP. Such a suppression gets weaker towards more peripheral collisions. The  $R_{CP}$ for the $D^0$ meson can be described by theoretical model calculations \cite{TheoryD0RAA}.  Also reported at the conference were the results of  $D^\pm$, $D^{*\pm}$ and $D_s$ in Au+Au collisions at $\sqrt{s_{NN}}=200$ GeV. A conclusion from these studies is that the total charm quark production cross-section per binary nucleon-nucleon collision is consistent between Au+Au and p+p collisions, but the relative abundances of different charm hadrons are different \cite{STARLc}.

Quarkonium production may be suppressed in heavy-ion collisions by the color-screening effect of the heavy quark-antiquark potential from unconfined quarks and gluons of the QGP. In particular, the $\Upsilon$ mesons, which are less affected by the recombination effect than $J/\psi$, are suggested as a good probe to study the thermodynamic properties of the QGP. STAR has updated the results on the $R_{AA}$ for the  $\Upsilon$ mesons in Au+Au collisions at $\sqrt{s_{NN}}=200$ GeV, as shown in Fig.~\ref{fig:Upsilon} \cite{Upsilon}. These results are obtained by combining the di-electron measurement from 2011 data and the di-muon measurement from 2014 and 2016 data. The results indicate that both the $\Upsilon$ ground and excited states are suppressed in central Au+Au collisions, and that the suppression for the excited states is stronger than that of the ground state. The latter is consistent with a naive expectation based on sequential melting of $\Upsilon$ states due to their different binding energies. 

\section{Chirality, Vorticity and Polarization Effects}
\begin{figure}
\includegraphics[width=0.45\textwidth]{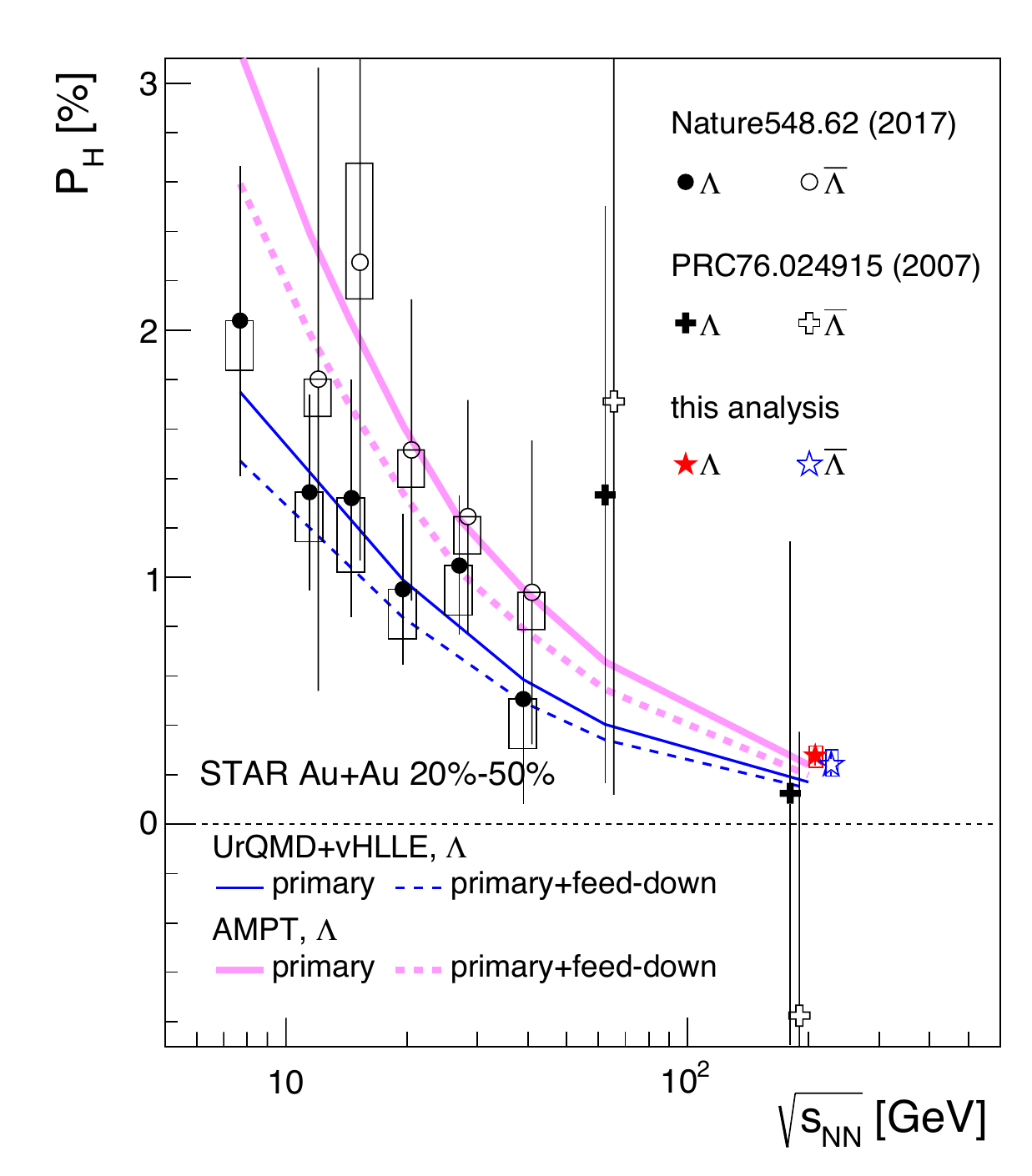}
\includegraphics[width=0.55\textwidth]{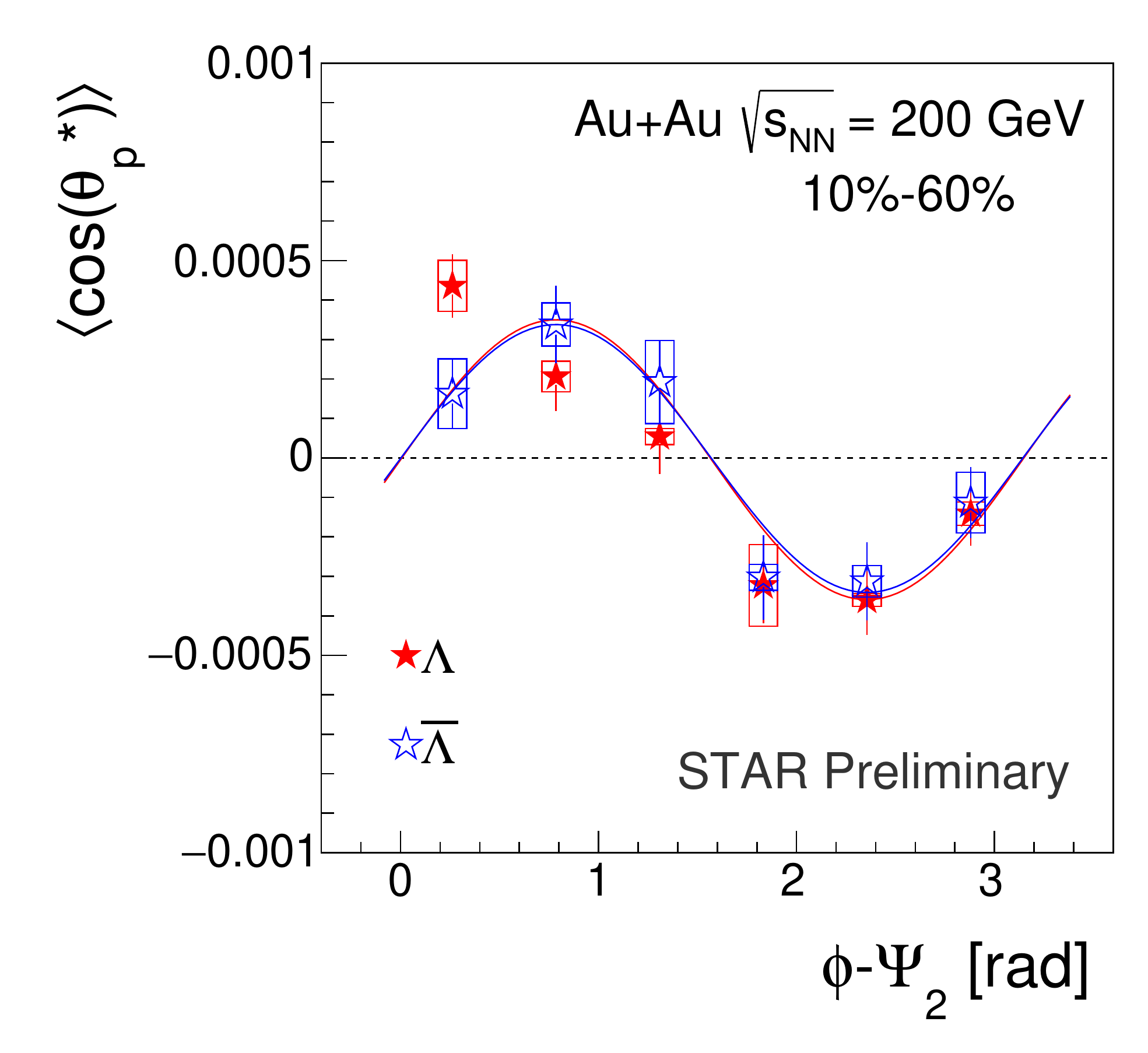}
\vspace{-0.8cm}
\caption{Left: global polarization for $\Lambda$ (solid points) and $\bar{\Lambda}$ (open points) in Au+Au collisions as a function of $\sqrt{s_{NN}}$. Right: local polarization for $\Lambda$ (solid points) and $\bar{\Lambda}$ (open points) as a function of azimuthal angle $\phi$ with respect to the second order reaction plane $\Psi_2$ in Au+Au collisions at  $\sqrt{s_{NN}}=200$ GeV. Vertical bars (open boxes) represent statistical (systematic) uncertainties.}
\label{fig:LambdaPol}

\vspace{0.3cm}
\includegraphics[width=0.46\textwidth]{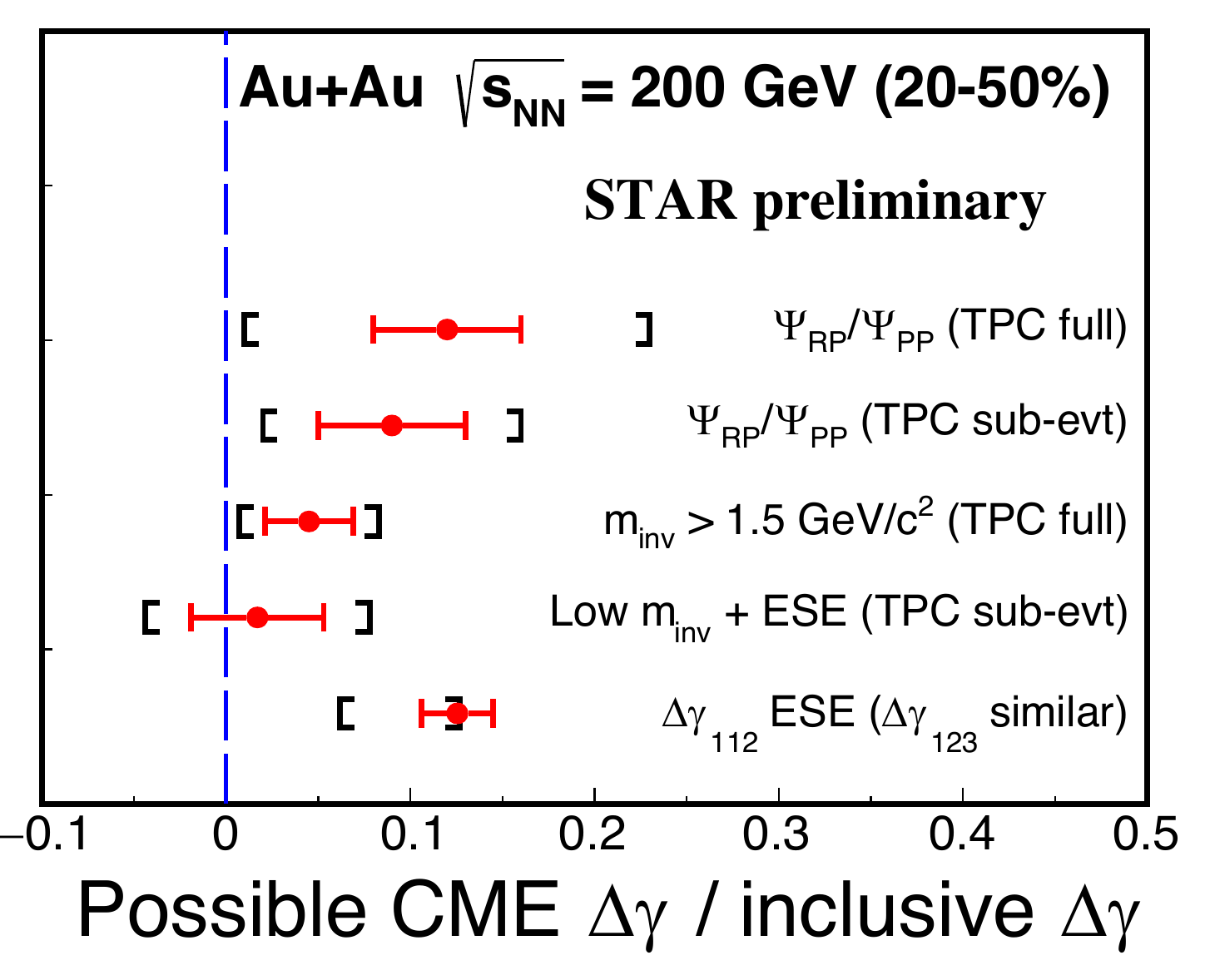}
\includegraphics[width=0.54\textwidth]{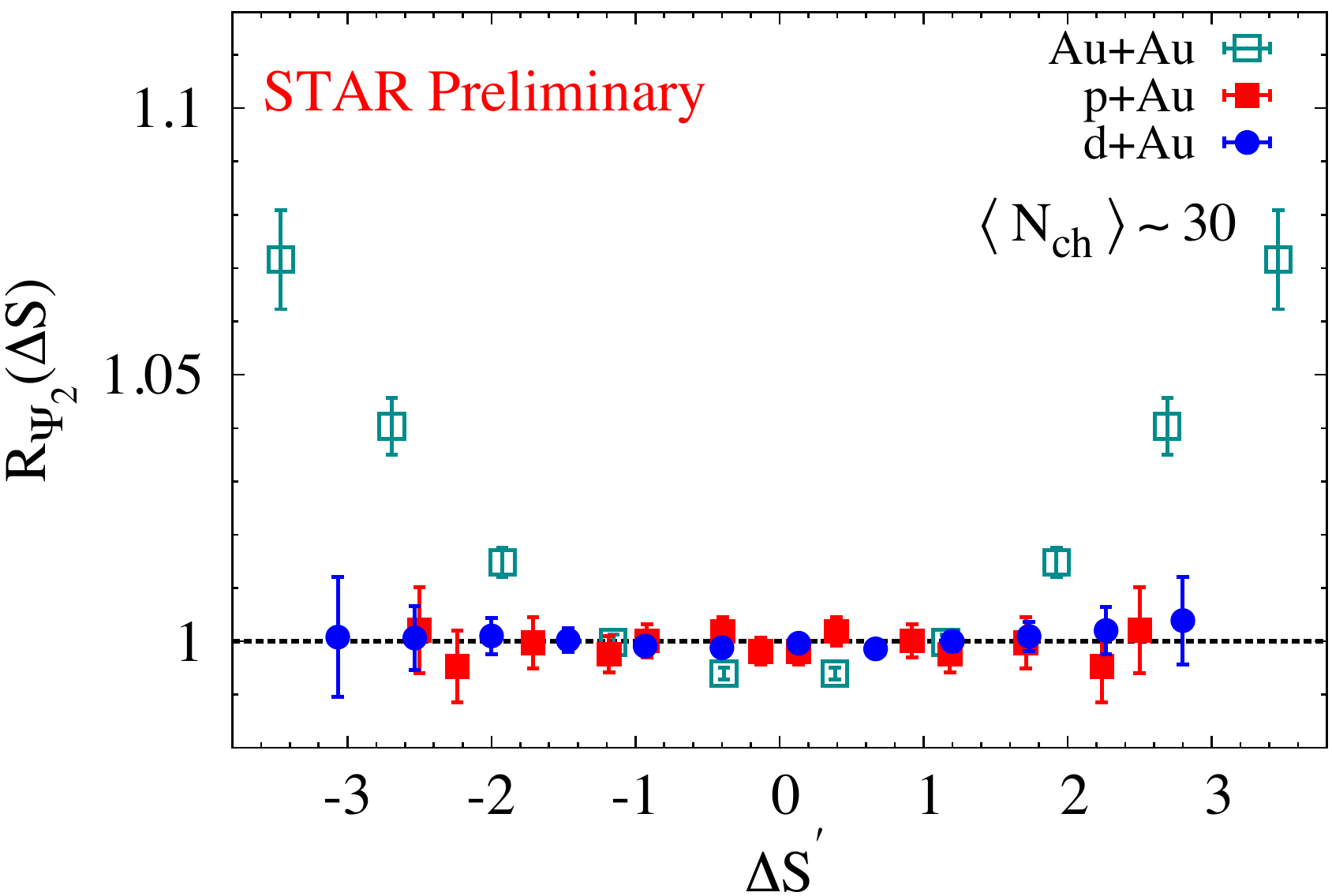}
\vspace{-0.8cm}
\caption{Left: relative contribution from possible CME signal to the measured $\Delta\gamma$. Horizontal bars (brackets) represent statistical (systematic) uncertainties. Right: $\Delta S$ distribution in p+Au (filled boxes), d+Au (filled circles) and Au+Au (open boxes) collisions at $\sqrt{s_{NN}}=200$ GeV with an average number of charged particles $\left<N_{ch}\right>\approx30$. Vertical bars represent statistical uncertainties.}
\label{fig:CME}

\center
\includegraphics[width=0.5\textwidth]{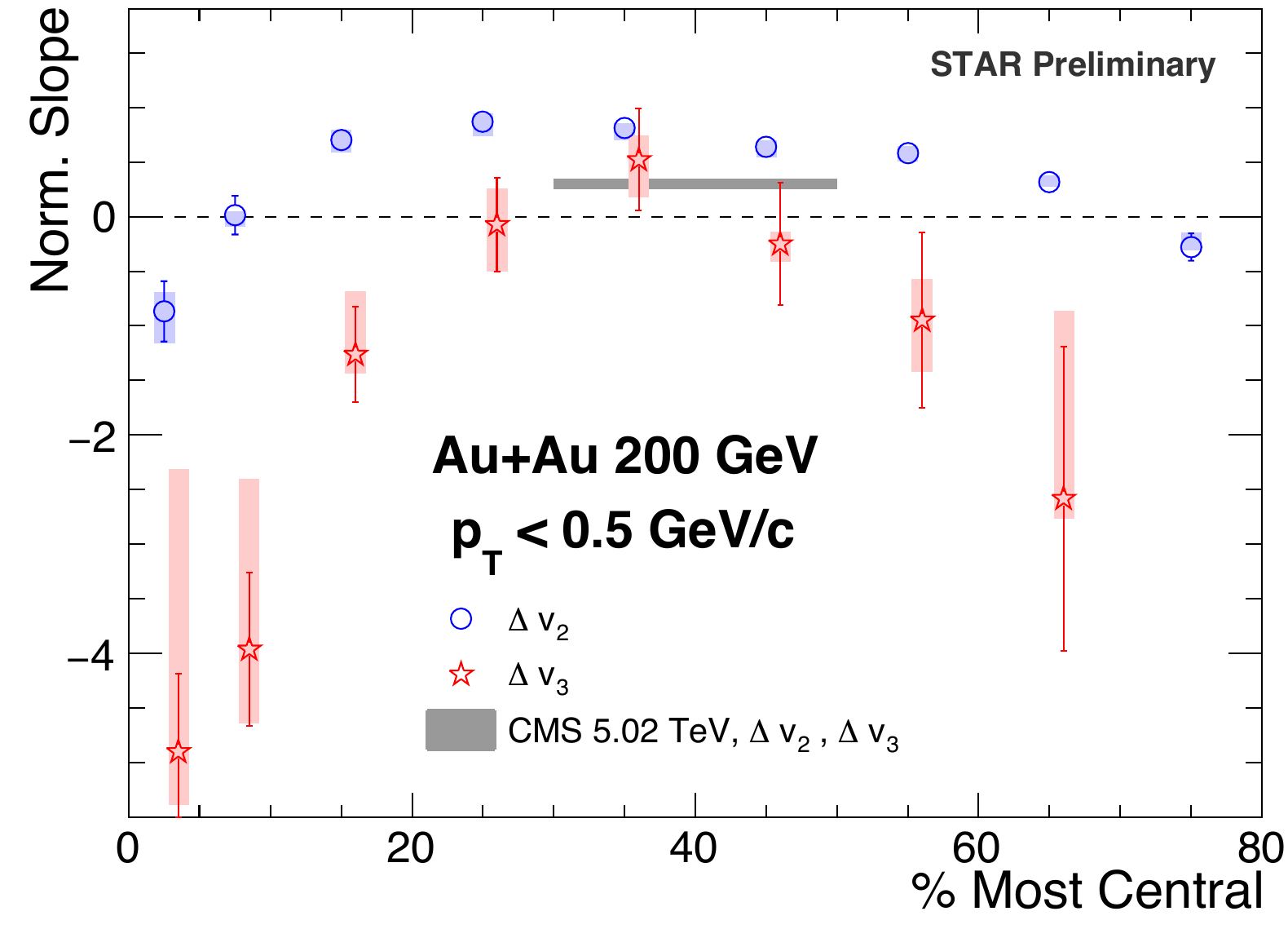}
\vspace{-0.4cm}
\caption{Difference in the self-normalized $v_2$ (open circles) and $v_3$ (stars) between positively and negatively charged particles as a function of centrality in Au+Au collisions at $\sqrt{s_{NN}}=200$ GeV. Also shown are results from CMS in Pb+Pb collisions at $\sqrt{s_{NN}}=5.02$ TeV (horizontal shaded area). Vertical bars (shaded areas) represent statistical (systematic) uncertainties.}
\label{fig:CMW}
\end{figure}

STAR reported the first observation of a $\Lambda$ global polarization with respect to the reaction plane in heavy-ion collisions, suggesting a strong vortical field in the QGP. This observation was made in Au+Au collisions at $\sqrt{s_{NN}}=7.7$-39 GeV, while the results at the RHIC top energies ($\sqrt{s_{NN}}=62$ and 200 GeV) were still limited by statistical uncertainty \cite{LambdaPolSTAR1}. From a much larger data sample than that in the previous study, the $\Lambda$ global polarization has also been observed at $\sqrt{s_{NN}}=200$ GeV, as shown in the left panel of Fig.~\ref{fig:LambdaPol} \cite{LambdaPolSTAR2}. The measured polarization value follows the trend suggested by lower energy results and is consistent with theoretical model calculations. At the QM2018 conference, STAR also reported the first observation of a $\Lambda$ local polarization along the beam direction with a quadrupole structure, as can be seen in the right panel of Fig.~\ref{fig:LambdaPol} \cite{LambdaPolSTAR3}. Such a structure could be generated by the presence of elliptic flow \cite{LambdaLocalPol}.  

Local parity-odd domains are theorized to form inside the QGP and manifest themselves as charge separation along the magnetic field axis via the Chiral Magnetic Effect (CME). A widely used observable to study the CME is $\Delta\gamma$, defined as the difference in the three-point correlation between unlike-sign and like-sign charged particle pairs. It is essential to know how much backgrounds contribute to the measured $\Delta\gamma$ value. Different methods have been used to separate the possible CME signal and background contributions in $\Delta\gamma$ at STAR. These methods consistently point to a large background contribution to $\Delta\gamma$ as can been seen in the left panel of Fig.~\ref{fig:CME} \cite{CMESTAR1}. A new observable, $\Delta S$ \cite{DeltaS}, has also been used to study the CME~\cite{CMESTAR2}. As can be seen in the right panel of Fig.~\ref{fig:CME}, the shape of the $\Delta S$ distributions is different between small collision systems (p+Au, d+Au)  and peripheral Au+Au collisions with similar number of charged particles. It is still under discussion how to interpret this result in terms of the CME. STAR took data in isobar (Ru+Ru and Zr+Zr) collisions in 2018 and is carrying  blind analyses of these data for the CME study.

The CME, if present, can intertwine with the Chiral Separation Effect (CSE)  to generate a finite electric quadrupole moment of the collision system, the so-called Chiral Magnetic Wave (CMW). STAR has carried a new study for the CMW, in which the difference in the elliptic flow between $\pi^+$ and $\pi^-$ ($\Delta v_2=v_2^{\pi^-}-v_2^{\pi^+}$) is measured as a function of the difference between the number of $\pi^+$ and $\pi^-$ ($A_{ch}=\frac{N_{\pi^+}-N_{\pi^-}}{N_{\pi^+}+N_{\pi^-}}$). It is found that the slope parameter $r$, obtained from fitting $\Delta v_2$ versus $A_{ch}$ to a linear function, is consistent with zero within uncertainty in small collision systems, and that $r$ is positive and larger in semi-central U+U collisions than that in semi-central Au+Au collisions \cite{CMWSTAR}. These observations resemble the expectation from the CMW. Another measurement involves the self-normalized difference in $v_n$ between positively and negatively charged particles, i.e., $\frac{2(v_n^--v_n^+)}{v_n^-+v_n^+}$. As can be seen in Fig.~\ref{fig:CMW}, such a difference is not the same between $v_2$ and $v_3$ in central and peripheral Au+Au collisions. This can not be explained by local charge conservation alone and the CMW remains as a viable interpretation of RHIC data.

\section{Collectivity}
\begin{figure}[b!]
\vspace{-0.2cm}
\center
\includegraphics[width=0.45\textwidth]{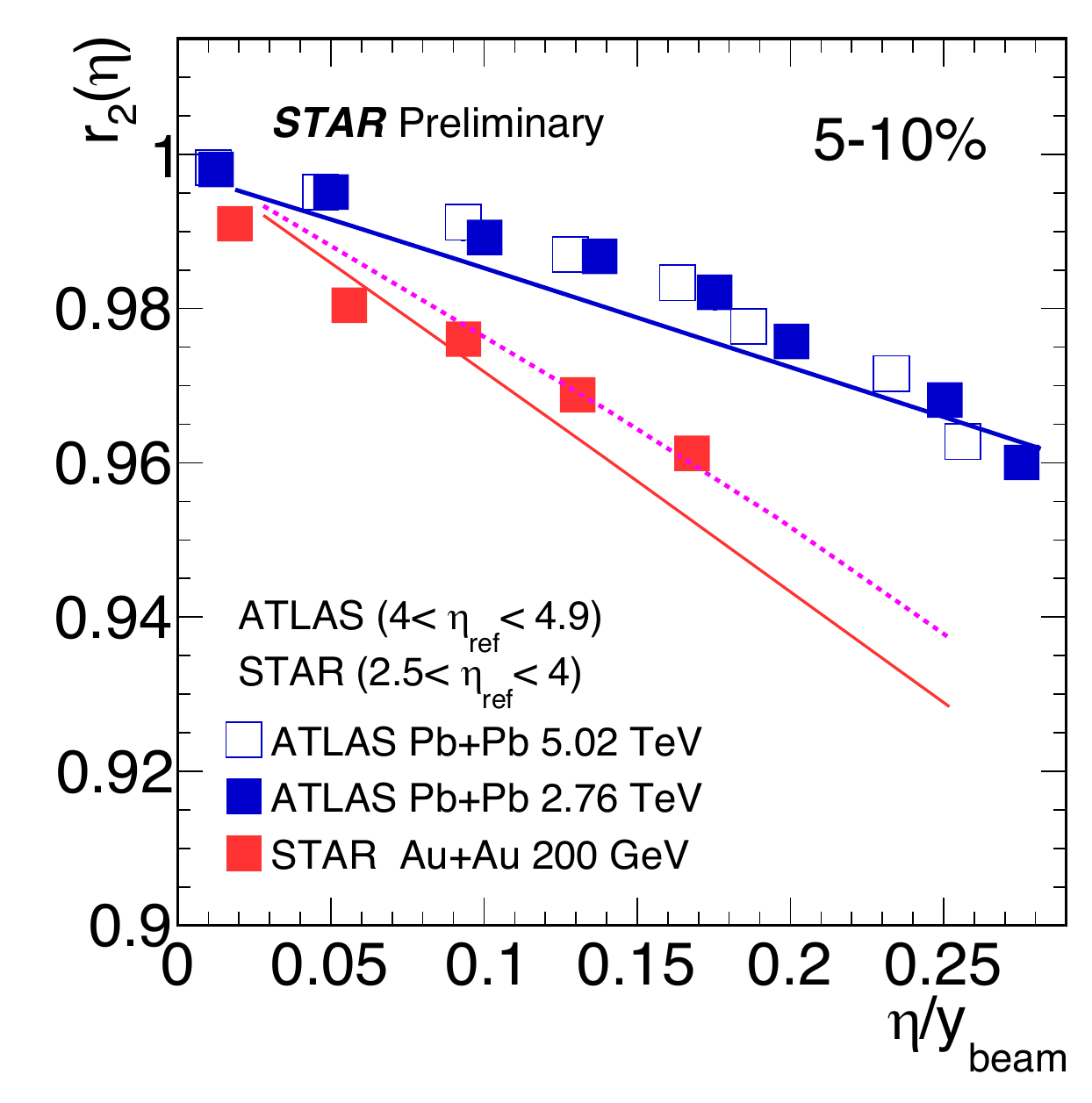}
\includegraphics[width=0.45\textwidth]{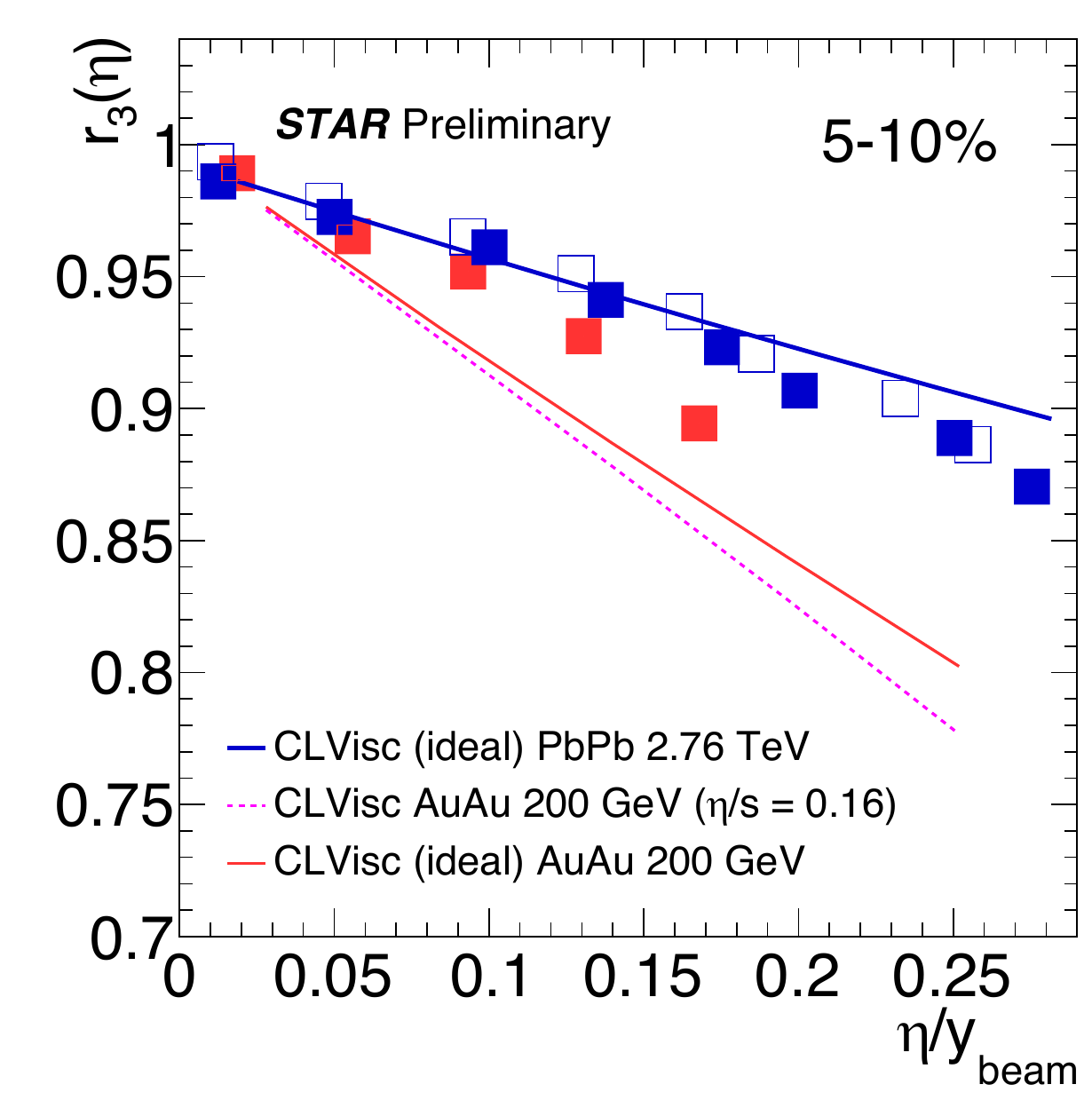}
\vspace{-0.6cm}
\caption{The factorization ratio $r_n$ for $v_2$ (left) and $v_3$ (right) in 5-10\% Au+Au collisions at $\sqrt{s_{NN}}=200$ GeV (red filled boxes), Pb+Pb collisions at 2.76 and 5.02 TeV (blue filled and open boxes). Also shown are (3+1)D hydrodynamic calculations \cite{DecorrelationTheory}.}
\label{fig:decorrelation}

\vspace{0.3cm}
\includegraphics[width=0.5\textwidth]{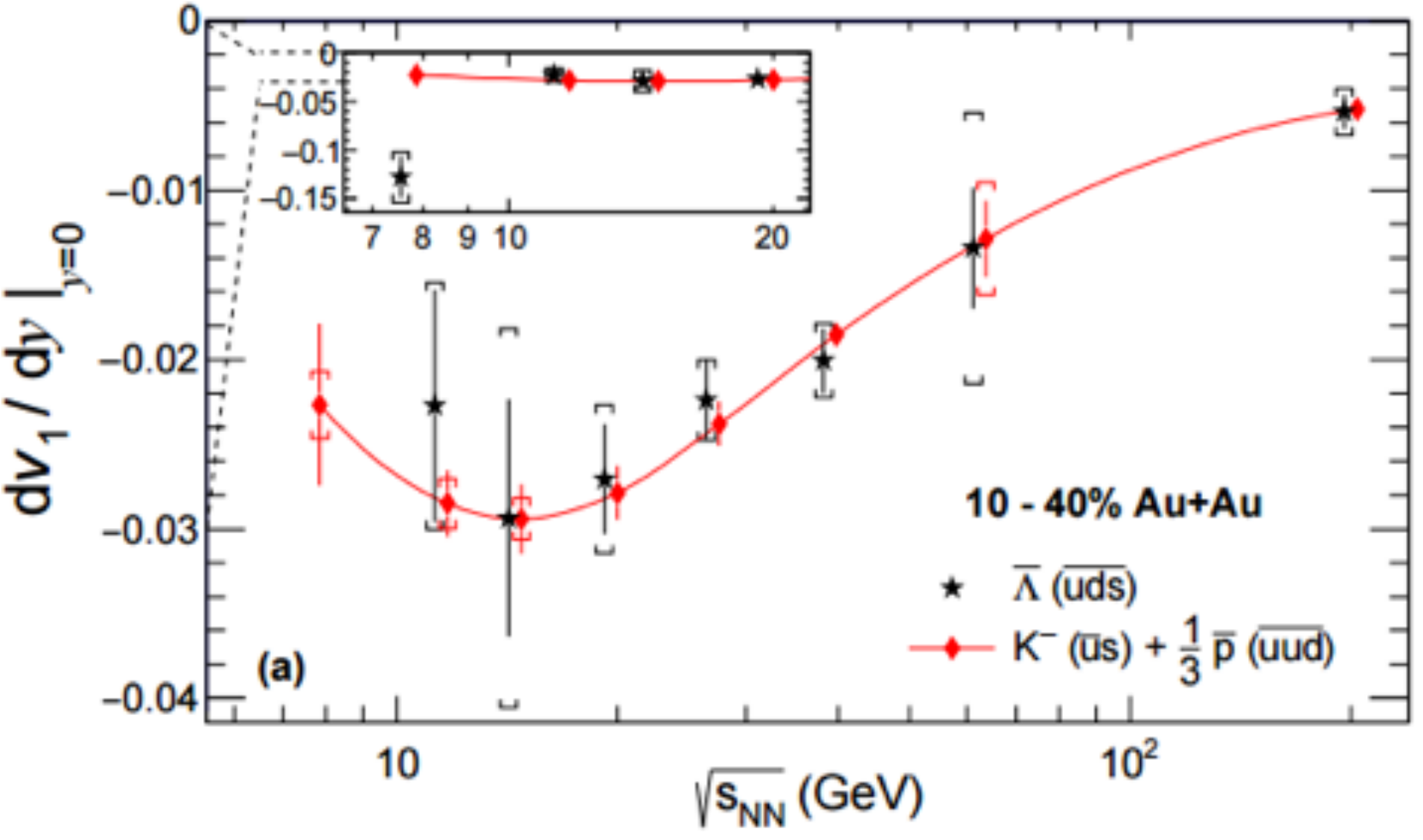}
\includegraphics[width=0.455\textwidth]{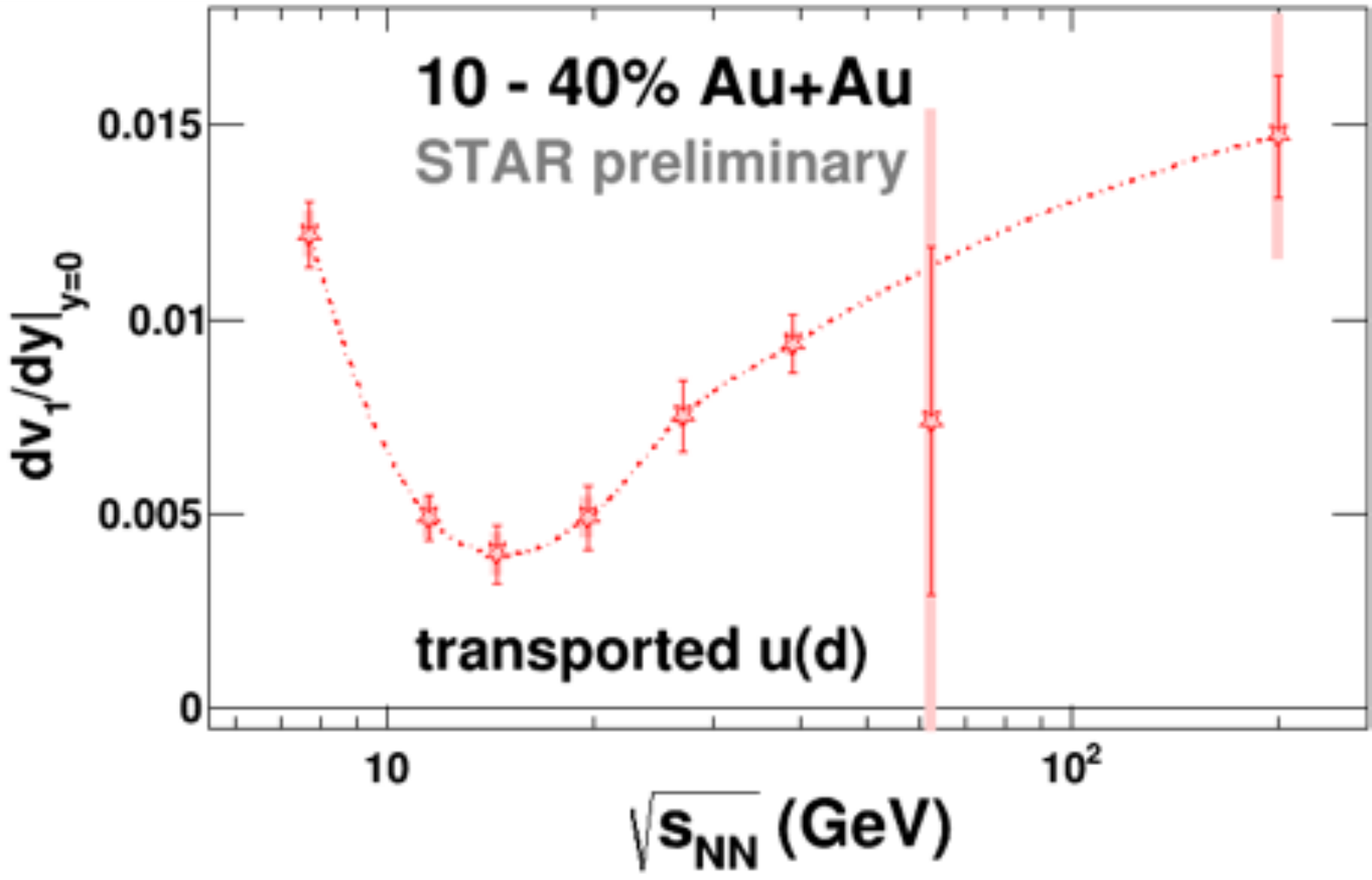}
\vspace{-0.3cm}
\caption{Left: directed flow slopes at mid-rapidity in Au+Au collisions at $\sqrt{s_{NN}}=7.7$-200 GeV to test the coalescence sum rule of produced quarks. Right: transported $u$ (and $d$) quark $v_1$ slope extracted from the STAR data assuming constituent quark coalescence. Vertical bars (brackets or shaded areas) represent statistical (systematic) uncertainties. }
\label{fig:v1}
\end{figure}

Longitudinal decorrelations of flow harmonics are sensitive to event-by-event fluctuations of the initial-collision geometry and final-state collective dynamics. Using the FMS as the reference detector, STAR has measured longitudinal decorrelations from 2016 data in Au+Au collisions at $\sqrt{s_{NN}}=200$ GeV \cite{DecorrelationSTAR}. The factorization ratios $r_2$ and $r_3$ \cite{Decorrelation} are used to measure the decorrelation of $v_2$ and $v_3$, respectively, between $\eta$ and $-\eta$ in the pseudo-rapidity range $|\eta|<1$, with respect to a common reference $2.5<\eta_{ref}<4$. They are found to exhibit a stronger decrease with the normalized rapidity ($\eta/y_{beam}$) than those at the LHC \cite{DecorrelationLHC}, as can be seen in Fig.~\ref{fig:decorrelation}. This is qualitatively consistent with (3+1)D hydrodynamic calculations \cite{DecorrelationTheory}. However, these calculations can not simultaneously describe both the RHIC and LHC data. Analyses of data in Au+Au collisions at lower energies are underway to provide more insight into the decorrelations.

The slope of the rapidity-odd $v_1$ at mid-rapidity ($dv_1/dy|_{y=0}$) is sensitive to the equation of state of the system. STAR has measured the $v_1$ slopes for 10 particle species ($\pi^\pm$, $p$, $\bar{p}$, $\Lambda$, $\bar{\Lambda}$, $K^\pm$ and $K_S$) at $\sqrt{s_{NN}}=7.7$-200 GeV \cite{v1STAR1}. The results are used to test the coalescence sum rule, and it is found that the sum rule holds for produced quarks at $\sqrt{s_{NN}}\geq11.5$ GeV, as can be seen in the left panel of Fig.~\ref{fig:v1}. Furthermore, the transported $u$ (and $d$) quark $v_1$ slope has also been estimated from these data by assuming constituent quark coalescence. As can be seen in the right panel of Fig.~\ref{fig:v1} \cite{v1STAR2}, the transported quark $v_1$ exhibits a minimum around $\sqrt{s_{NN}}=14.5$ GeV, which could be related to softening of the equation of state of the system.

\section{Hypertriton Binding Energy and Particle-Antiparticle Mass Difference}
STAR  has recently published a paper on the hypertriton ($^3_\Lambda H$) lifetime \cite{HypertritonSTAR1}. Using the 2014 and 2016 data with the HFT, the hypertriton signal-to-background ratio is significantly improved, allowing precise determination of the hypertriton binding energy and mass difference between the hypertriton ($^3_\Lambda H$) and anti-hypertriton ($^3_{\bar{\Lambda}}\bar{H}$) \cite{HypertritonSTAR2}. These results could have a large impact on understanding the hyperon-nucleon (Y-N) interaction and the CPT symmetry in the hypernucleus sector. 


\section{Summary and Outlook}
Many new results from the RHIC top energies, beam energy scan and fixed-target data taken before 2017 were reported by STAR at the QM2018 conference. In 2018 STAR took data in Ru+Ru and Zr+Zr collisions to study the CME, and Au+Au collisions at $\sqrt{s_{NN}}=27$ GeV to look for difference in the global polarization between $\Lambda$ and $\bar{\Lambda}$ due to the initial magnetic field. In 2019-2021, STAR will take data in both collider and fixed-target modes at $\sqrt{s_{NN}}=3.0$-19.6 GeV to look for signatures of the critical point and phase transition of the QCD phase diagram, with new detectors including inner TPC, endcap TOF, and event plane detector for improved acceptance, particle identification and event plane reconstruction \cite{UpgradeSTAR}. STAR is also working on R\&D for new forward calorimeter and tracking systems ($2.5<\eta<4$) for detailed studies of the initial conditions, longitudinal decorrelation and rapidity dependence of $\Lambda$ global polarization in A+A collisions, and the partonic and spin structure of the nucleon and nuclei in p+p and p+A collisions in 2021+. 

\vspace{0.2cm}

\textbf{Acknowledgement} This work was partly supported by the U.S. Department of Energy (Grant No. DE-FG02-94ER40865).





\bibliographystyle{elsarticle-num}
\bibliography{<your-bib-database>}

\begin{thebibliography}{00}
\bibitem{spinalignment} C.~Zhou (for the STAR collaboration), contribution to these proceedings.
\bibitem{cumulant} T.~Nonaka (for the STAR collaboration), contribution to these proceedings.
\bibitem{flow} N.~Magdy (for the STAR collaboration), contribution to these proceedings.
\bibitem{small} S.~Huang (for the STAR collaboration), contribution to these proceedings.
\bibitem{femtoscopy} S.~Siejka (for the STAR collaboration), contribution to these proceedings.
\bibitem{jet} K.~Jiang (for the STAR collaboration), contribution to these proceedings.
\bibitem{dihadron} R.~Aoyama (for the STAR collaboration), contribution to these proceedings.
\bibitem{FXT} Y.~Wu (for the STAR collaboration), contribution to these proceedings.
\bibitem{D0v1Theory} S.~Chatterjee and P.~Bozek, Phys.~Rev.~Lett.~120, 192301 (2018); S. Das et al., Phys. Lett. B 768, 260 (2017).
\bibitem{D0v1STAR} S.~Singha (for the STAR collaboration), contribution to these proceedings.
\bibitem{STARLc} S.~Radhakrishnan (for the STAR collaboration, contribution to these proceedings.
\bibitem{ALICELc} ALICE Collaboration, JHEP 04, 108 (2018).
\bibitem{TheoryLc} S.H.~Lee et al., Phys.~Rev.~Lett.~100, 222301 (2008); S.~Ghosh et al., Phys.~Rev.~D 90, 054018 (2014). 
\bibitem{PYTHIA} T.~Sjstrand, S.~Mrenna and P.~Skands, Journal of High Energy Physics 2006, 026 (2006).
\bibitem{TheoryD0RAA} Y.~Xu et al., Phys.~Rev.~C 97, 014907 (2018); S.~Cao et al., Phys.~Rev.~C 94, 014909 (2016).
\bibitem{Upsilon} P.~Wang (for the STAR collaboration), contribution to these proceedings.
\bibitem{UpsilonRAACMS} CMS collaboration, Phys.~Lett.~B 770, 357 (2017).
\bibitem{LambdaPolSTAR1} STAR collaboration, Nature 548 (2017) 62;  Phys. Rev. C 76, 024915 (2007).
\bibitem{LambdaPolSTAR2} STAR collaboration, Phys.~Rev.~C 98 014910 (2018).
\bibitem{LambdaPolSTAR3} T.~Niida (for the STAR collaboration), contribution to these proceedings.
\bibitem{LambdaLocalPol} F.~Becattini and I.~Karpenko, Phys.~Rev.~Lett.~120, 012302 (2018); S.~Voloshin, EPJ Web Conf.~17 10700 (2018).
\bibitem{CMESTAR1} J.~Zhao (for the STAR collaboration), contribution to these proceedings.
\bibitem{DeltaS} N.~Magdy et al., Phys.~Rev.~C 97, 061901 (2018). 
\bibitem{CMESTAR2} N.~Magdy (for the STAR collaboration), poster presentation at the QM2018 conference.
\bibitem{CMWSTAR} Q.-Y.~Shou (for the STAR collaboration), contribution to these proceedings.
\bibitem{DecorrelationSTAR} M.~Nie (for the STAR collaboration), contribution to these proceedings.
\bibitem{Decorrelation} V.~Khachatryan, et al., Phys.~Rev.~C 92, 034911 (2015).
\bibitem{DecorrelationLHC} ATLAS collaboration, Eur.~Phys.~J.~C 78, 142 (2018).
\bibitem{DecorrelationTheory} L.-G.~Pang et al., Phys. Rev. D 91, 074207 (2015); L.-G.~Pang, H.~Petersen, X.-N.~Wang, Phys. Rev. C 97, 064918 (2018).
\bibitem{v1STAR1} STAR collaboration, Phys.~Rev.~Lett.~112, 162301 (2014); Phys.~Rev.~Lett.~120, 062301 (2018).
\bibitem{v1STAR2} G.~Wang (for the STAR collaboration), contribution to these proceedings. 
\bibitem{HypertritonSTAR1} STAR collaboration, Phys.~Rev.~C 97, 054908 (2018).
\bibitem{HypertritonSTAR2} P.~Liu (for the STAR collaboration), contribution to these proceedings.
\bibitem{UpgradeSTAR} Q.~Yang (for the STAR collaboration), contribution to these proceedings.

\end{thebibliography}



\end{document}